\documentclass[final,1p,times]{elsarticle} 

\usepackage{amssymb}

\newcommand {\bc}{\begin{center}}
\newcommand {\ec}{\end{center}}
\def\lsim{\mathrel{\rlap{\lower4pt\hbox{$\sim$}}
    \raise1pt\hbox{$<$}}}               
\def\gsim{\mathrel{\rlap{\lower4pt\hbox{$\sim$}}
    \raise1pt\hbox{$>$}}}  
\newcommand {\bea}{\begin{eqnarray}}
\newcommand {\eea}{\end{eqnarray}}
\newcommand {\be}{\begin{equation}}
\newcommand {\ee}{\end{equation}}
\newcommand{\xvect}{{\bf x}_{\bot}}

\newcommand{\kvect}{{\bf k}_{\bot}}

\usepackage{graphicx}

\journal{Nuclear Physics A}

\begin{document}

\begin{frontmatter}



\title{Thermalization of the world's smallest fluids: recent developments}


\author{Raju Venugopalan}

\address{Bldg. 510A, Physics Department, Brookhaven National Lab, Upton, NY, 11973, USA}

\begin{abstract}
The late Gerry Brown was not shy to tackle complex scientific problems that took time to play out but yielded in the end a deeper understanding of many-body phenomena. In this note, prepared for a memorial volume in his honor, we provide a  perspective on a couple of outstanding scientific puzzles that have their origin in our understanding of the thermalization of matter in ultrarelativistic heavy ion collisions, and possibly, in high multiplicity proton-proton and proton-nucleus collisions. 
\end{abstract}




\end{frontmatter}


\section{Introduction}

The successful comparison of hydrodynamical models to a wide range of data from heavy ion collisions suggests that the produced quark-gluon fluid is a viscous fluid with perhaps the lowest known viscosity to entropy density ratio ($\eta/s$) in nature. The low values of $\eta/s$, coupled with the fact that these hydrodynamical models employ an equation of state, appear to indicate that the matter is thermal, or at least nearly isotropic, with the ratio of the longitudinal pressure to the transverse pressure close to unity. Further, to reproduce key features of the data, it appears important that hydrodynamics be applicable at very early times of less than a Fermi after the collision. 

There is some elasticity to the above conclusions, and it is conceivable that the hydrodynamic paradigm may be modified. Nevertheless, the phenomenology is sufficiently robust to approach seriously. From a theoretical perspective, at first glance, it seems astonishing that hydrodynamics is applicable at all to such small systems, and at such early times. Hydrodynamics is an excellent effective field theory of QCD, but for long wavelength modes and at late times~\cite{Arnold:1997gh}.  In kinetic theory frameworks, the scattering rates of quarks and gluons have to be sufficiently strong to counter the rapid longitudinal expansion of the system. This appears challenging. From these elementary, and perhaps naive considerations, to paraphrase a quote by Wigner in another context, hydrodynamics is ``unreasonably effective".  

A weak link in hydrodynamic models is the statement that the system isotropizes/thermalizes at very early times.  Most hydrodynamic models choose an initial time $\tau_i = 0.4-0.6$ fm. Nearly all these models ignore the pre-equilibrium dynamics prior to this time. The one model that does incorporate pre-equilibrium dynamics, the IP-Glasma model, does so imperfectly~\cite{STV}. There is therefore a practical problem of how and when pre-equilibrium dynamics can be matched on to a hydrodynamic description. This of course is tied to resolving the more general conceptual problem of how thermalization occurs in QCD. The latter will be one of the two subjects of the discussion here. 

Even if thermalization does occur in heavy ion collisions, as the hydrodynamic models suggest, there is the interesting question of whether this framework is applicable to smaller size systems. How and where does hydrodynamics break down? Does it apply to p+A and p+p collisions, as some interpretations of data (on long range rapidity correlations) in these collisions suggest? If it works for high multiplicities, at what multiplicities do we see an onset of hydrodynamic behavior? Are there alternative explanations for what is seen in the data? The interpretation of long range rapidity correlations in p+p and p+A collisions will be the other topic discussed here. 

Both topics will be discussed within weak coupling frameworks here. It is a common misunderstanding that weak coupling implies weakly interacting. That is not the case for systems with high occupancy. It is a legitimate question to ask whether weak coupling is the right framework for heavy ion collisions--at RHIC and LHC, the coupling is not particularly weak. At one level, an answer is that this is the only framework we know how to compute in systematically and reliably. But this answer is also profoundly unsatisfying. A better answer is that weak coupling frameworks describe many non-trivial features of heavy ion collisions. It is however not a universal panacea, which disappoints some people, but that can't be helped--until some smart person solves QCD.

The lack of a satisfactory framework to address dynamical aspects of QCD in strong coupling is a powerful motivation for AdS/CFT duality inspired frameworks. The questions regarding the applicability of these methods to heavy ion collisions are well known, and I will not revisit them here. The next section will discuss the problem of thermalization in weak coupling. We will then discuss the recent results from p+p and p+A collisions on collimated long range rapidity correlations (``the ridge"). 
Since many of the issues discussed are open, and are the subject of much debate, conclusions may be premature. I will conclude instead with some personal reminiscences of Gerry Brown, whose early mentorship made it possible for me, however imperfectly, to tackle these issues.

\section{A weak coupling treatment of the thermalization process in QCD}

Multiparticle production at central rapidities is dominated by gluon configurations carrying small fractions $x$ of the momenta of  the colliding nuclei. Perturbative QCD (pQCD) predicts, and data from HERA confirm, that the occupancy of small $x$ modes in a proton is large for fixed momentum transfer ${\bf Q}^2$. The occupancy saturates at $1/\alpha_S$ for any given ${\bf Q}^2\gg\Lambda_{\rm QCD}^2$, for sufficiently small $x$. This phenomenon is called gluon saturation, and the saturation scale $Q_S (x)$ is the dynamical (semi-) hard scale at which maximal occupancy is attained for a given $x$~\cite{Gribov:1984tu,Mueller:1985wy}. A small $x$ probe with ${\bf Q}^2 < Q_S^2$, will see a nearly perfectly absorptive black disk of gluons. 

The Color Glass Condensate (CGC) is a classical effective field theory of such highly occupied gluon configurations~\cite{CGC}. Systematic computations (LO, NLO,...) are feasible in this framework. An ever present issue is factorization: what gluon modes are universal--resolved when the nuclei are coarse grained at different resolutions--and what modes participate in the scattering process under consideration. Factorization in nucleus-nucleus collisions has been proven to leading logarithmic accuracy; in plainspeak, the largest logarithmically divergent terms in $x$, at each order in perturbation theory, can be resummed into nuclear wavefunctionals that can be probed through various final states. 

In nucleus-nucleus collisions, the leading order CGC  result for components of the stress energy tensor $T^{\mu\nu}$ is given by solutions of Yang-Mills equations for two lightlike classical sources (representing the large $x$ static color charges of the two nuclei) with null initial conditions for fields at negative infinity~\cite{MV,CYM}. The numerical solution to this problem is known~\cite{CYM-numerics}. On the (proper) time scale $\tau\sim 1/Q_S$, the system is highly anisotropic: $P_T^x = P_T^y=\varepsilon/2$ and $P_L=0$, where $\varepsilon$ is the energy density, $P_T^{x,y}$ the transverse pressures in the $x$ and $y$ directions orthogonal to the beam, and $P_L$ is the longitudinal pressure of the expanding nuclei after the collision. This is straightforwardly understood to be a consequence of the classical field solutions $A_{\rm class.}^\mu \equiv A_{\rm class.}^\mu(x_\perp,\tau)$ being independent of the space-time rapidity $\eta$. 

This non-equilibrium matter, often called the Glasma~\cite{Glasma}, is far from isotropy\footnote{In what follows, we will replace $Q_S$ by $Q$; the former is the saturation scale in the nuclear wavefunctions, the latter is the hard scale in the Glasma. While the latter derives from the former, they do not have to be identical.}  at $\tau\sim 1/Q$. The large $p_T > Q$ modes are frozen out, with distributions given by leading order pQCD expressions. Interestingly, the softer $p_T$ modes are described by a 2-D classical Bose-distribution $\sim Q/p_T$~\cite{CYM-numerics2}.  What happens at NLO ?  Parametrically small quantum fluctuations ( O(1) relative to the leading $1/g$ classical field) can grow exponentially as $\exp(\sqrt{Q \tau})$~\cite{Romatschke:2005pm}--this is due to an instability of the small fluctuation propagator in the classical field, an instability that has been interpreted previously in the language of Weibel~\cite{Weibel} and Nielsen-Olesen~\cite{Nielsen-Olesen} instabilities. So, if we were to try and do a systematic computation, 
$T^{\mu\nu} = T_{\rm LO}^{\mu\nu} + T_{\rm NLO}^{\mu\nu} + \cdots$, we would find terms at NLO that would be as large as the LO contribution at times $Q \tau_{\rm inst.} \sim \ln(\alpha_S^{-2})$, indicating a breakdown of the perturbative expansion at these times. Note that this power counting makes sense only in very weak coupling; for realistic values of $\alpha_S$ relevant for experiments, the $\ln^2(1/\alpha_S^2)$ is not appreciably different from $Q \tau \sim 1$. 

Just as in the case of small $x$ resummations, the leading $(g^2 \exp(\sqrt{Q \tau})^n$ at each nth sub-leading order in perturbation theory can be resummed and written as~\cite{Gelis:2007kn,Dusling:2011rz}
\begin{equation}
T_{\rm resum.}^{\mu\nu} (x) = \int [D\alpha]\;F_0 [\alpha] T_{\rm LO}^{\mu\nu} [A_{\rm class.} + \alpha](x)\, ,
\label{eq:EM-tensor}
\end{equation}
where 
\begin{equation}
F_0[\alpha] = \exp\left( - \frac{1}{2} \int_{\tau =0^+} d^3 u d^3 v \alpha(u) \Gamma^{-1}(u,v) \alpha (v)\right) \, ,
\end{equation}
denotes the spectrum of fluctuations, and $\Gamma(u,v)$, the small fluctuations propagator in the classical background field at proper time $\tau=0^+$ after the collision. To compute $\Gamma$, it is sufficient to compute solutions of the small fluctuation equations of motion (initialized as plane waves at negative infinity) through the field strengths of one of the nuclei and into the forward light cone at $\tau=0^+$ after the collision. This is quite challenging but precisely such a computation was performed recently~\cite{Epelbaum:2013waa}. 

Computing Eq.~(\ref{eq:EM-tensor}) is equivalent to solving the 3+1-D classical Yang-Mills equations with the initial condition 
\begin{equation}
{\bf A}_{\rm init.}^\mu = A_{\rm class.}^\mu + \int d\mu_{_K}\;\Big[c_{_K}\,a_{_K}^\mu(x)+c_{_K}^*\,a_{_K}^{\mu*}(x)\Big] \, .
\label{eq:quantum}
\end{equation}
The  coefficients $c_{_K}$, where $K$ collectively denote quantum numbers labeling the basis of solutions, are random Gaussian-distributed complex numbers. The explicit expressions for the small fluctuations and their conjugate momenta, denoted here by $a_K^\mu (x)$, are given in \cite{Epelbaum:2013waa}. 

While the computations in \cite{Epelbaum:2013waa} may be important for a variety of final states sensitive to very early times after the collision, their relevance for the problem of thermalization in heavy ion collisions is unclear. Firstly, the resummation in 
Eq.~(\ref{eq:EM-tensor}) will receive sub-leading corrections $g^p (g^2\exp(\sqrt{2 Q\tau}))^n$ with $p >0$. These all become of order unity on time scales that are parametrically $Q^{-1} \ln(1/\alpha_S)$; this suggests that the resummation procedure may be fragile when extended to later times. Secondly, it has been argued that an effective ``memory loss" of initial conditions may occur when oscillating functions, each corresponding to a Gaussian distributed initial condition of varying amplitude, are summed over~\cite{Berges:2004ce,Dusling:2010rm}. More generally, we argued that the `late-time' dynamics of the Glasma is described by a universal nonthermal fixed point, the properties of which are independent of the initial conditions~\cite{Berges:2013eia}. 

The picture we have developed is as follows. By timescales $Q\tau_0 \gg \ln^2(\alpha_S^{-1})$, phase decoherence sets in and the classical fields, in $A^\tau=0$ gauge,  can be decomposed into quasi-particle modes described by single particle distributions, as 
\begin{eqnarray}
\label{eq:quantum2}
A_{\mu}^a(\tau_0,\xvect,\eta)=\sum_{\lambda=1,2}\int \frac{d^2 \kvect}{(2\pi)^2}\, \frac{d \nu}{2\pi} \, \sqrt{f(\kvect,\nu,\tau_0)}\, \left[c_{\lambda,a}^{\kvect\nu}\, \xi^{(\lambda)\kvect\nu+}_{\mu}(\tau_0)\, e^{i \kvect\xvect}\,e^{i\nu\eta}+c.c. \right]\;, 
\end{eqnarray}
and a corresponding expression for the canonically conjugate momenta. Here $\xi^{(\lambda)\kvect\nu+}_{\mu,a}(\tau)$ denote the (time dependent) transverse polarization vectors of modes with transverse momentum $\kvect$, rapidity wave number $\nu$ and polarization index $\lambda=1,2$ in the non-interacting theory\footnote{For detailed expressions, see \cite{Berges:2013fga}.} and $c.c.$ denotes complex conjugation. The classical-statistical ensemble is defined by the distribution of the coefficients $c_{\lambda,a}^{\kvect\nu}$, which are the complex Gaussian random numbers introduced previously. The single particle distributions are parametrized as $f(p_{T},k_{z},\tau_0) = \frac{n_0}{8\pi\alpha_S}\, \Theta\!\left( Q - \sqrt{p_T^2+(\xi_0 p_z)^2}\right)$. If the dynamics is faithfully captured by quasi-particle excitations for $Q\tau_0 \gg \ln(\alpha_S^{-1})$, details of the evolution from earlier times $\tau \sim Q^{-1}$ up to $\tau_0$ are subsumed in the initial occupancy $n_0$, and in the anisotropy of the initial momentum distribution $\xi_0$. 

With the initial conditions of Eq.~(\ref{eq:quantum2}) for the gauge fields, and likewise, an initial condition for canonical conjugate momenta, the 3+1-D Yang-Mills equations can be solved for longitudinally expanding non-Abelian plasmas.  To ensure the dynamics is classical for the entire lifetime of the evolution, the amplitude of the gauge fields has to be chosen to be 
much larger than those obtained by replacing $f$ by the quantum `$1/2$'. For the results of the simulations discussed here, we have chosen $\alpha_S=10^{-5}$ corresponding to $Q\tau_0\sim 100$. We will comment shortly on the extrapolation of these results to more realistic values of $\alpha_S$. 

We find strikingly that the classical-statistical evolution of the system at late times demonstrates self-similar behavior independent of the initial conditions. Specifically, the single particle distributions extracted from the numerical simulations at later times by inverting Eq.~(\ref{eq:quantum2}), and the corresponding expression for canonically conjuage momenta, satisfy
\begin{eqnarray}
\label{eq:scalingf}
f(p_T,p_z,\tau)=(Q\tau)^{\alpha}f_S\Big((Q\tau)^\beta p_T,(Q\tau)^\gamma p_z\Big) ,
\end{eqnarray}
where $f_S$ denotes a {\em stationary} distribution in time, that describes the spectral properties of the non-thermal fixed point. This is seen for a wide range of $n_0$ and $\xi_0$ values, and has been checked to hold for initialization times ranging from $Q\tau_0 =20$ and $Q\tau_0=1000$. In each case, regardless of the initial condition, the system flows to the same attractor solution by $\tau/\tau_0 < 10$. 

The scaling exponents $\gamma$ and $\beta$ describe the temporal evolution of the characteristic hard longitudinal and transverse momentum scales respectively: 
\begin{equation}
\Lambda_L^2(\tau) \sim (Q \tau)^{-2 \gamma}\, , \quad \Lambda_T^2(\tau) \sim (Q \tau)^{-2 \beta} \, .
\label{eq:scales}
\end{equation}
One can therefore equivalently write $f(p_T,p_z,\tau)=(Q\tau)^{\alpha}f_S\Big( p_T/\Lambda_T, p_z/\Lambda_z\Big)$. The scaling exponent $\alpha$ describes the overall decrease of the distribution amplitude in time. The hard scales in Eq.~(\ref{eq:scales}) are gauge invariant quantities~\cite{Kurkela:2012hp} that can be measured on the lattice; $\beta$, $\gamma$ and the combination 
$\alpha-3\beta-\gamma$ can be extracted from the logarithmic derivatives of $\Lambda_T$, $\Lambda_L$, and the energy density respectively. Very careful measurements, on the largest lattices to date, have been performed and one extracts~\cite{Berges:2013fga} $\beta\sim 0$, $\gamma\sim 1/3$ and $\alpha\sim-2/3$. The scaling exponents can also be extracted from moments of the single particle distribution in Eq.~(\ref{eq:scalingf}); within systematic uncertainties, they are consistent with the values quoted here. 

The decrease in the occupancy as $\tau^{-2/3}$ implies that the system that began with an occupancy of $\alpha_S^{-1}$ will reach an occupancy of unity at time scales $Q\tau\sim \alpha_S^{-3/2}$. The classical simulations will no longer be valid after this time scale. However the parametrically large window of $\alpha_S^{-3/2}$, where these simulations are valid, provides insight into the rich dynamics of the thermalization process. The transverse momentum of hard gluons does not change during this time; their occupancy only decreases due to the redshift in the longitudinal momentum; they are  distributed as $T/p_T$, with $T=Q\tau^{-2/3}$, very early in the evolution. If the system were to free-stream, the ratio $P_L/P_T$ would decrease as $\tau^{-2}$; instead, the strong scattering of 
gluons prevents this, and $P_L/P_T \sim \tau^{-2/3}$ instead. While the anisotropy does increase, the lowest value of the ratio $P_L/P_T$ is $\alpha_S^{1/2}$ at the end of the classical regime. 

When occupancies are less than $\alpha_S^{-1}$, but greater than unity, there exists a dual description of the system in terms of 
either classical fields or kinetic theory~\cite{Mueller:2002gd,Jeon:2004dh}. The latter framework however is sensitive to dynamics at the Debye screening scale that is non-perturbative; this sensitivity allows for a number of kinetic scenarios, including 
the possible late time role of plasma instabilities~\cite{bib:KMII} or transient Bose-Einstein condensates~\cite{bib:BGLMV}. Instead, we find that our numerical simulations are consistent with the  late stage classical dynamics of the ``bottom-up" thermalization scenario~\cite{bib:BMSS}. In this kinetic scenario, the dynamics of overoccupied gluons is dominated by small angle elastic scattering. 

The self-similar scaling behavior extracted from both gauge invariant and gauge fixed observables in the classical-statistical field simulations of the expanding non-Abelian plasma finds a simple a posteriori explanation in the context of wave turbulence. Such analyses were performed previously in scalar field theories that are used to describe the thermalization process in the post-inflationary early universe~\cite{Micha:2004bv}. In close analogy to this earlier work, the  self-similar behavior characteristic of wave turbulence in the gauge theory can be interpreted in the framework of a kinetic equation
\begin{eqnarray}
\label{eq:Boltzmann}
\left[\partial_{\tau}-\frac{p_z}{\tau}\partial_{p_z}\right]f(p_T,p_z,\tau)=C[p_T,p_z,\tau;f] \;,
\end{eqnarray}
for the single particle distribution $f(p_T,p_z,\tau)$ with a generic collision term $C[p_T,p_z,\tau;f]$ for $n\leftrightarrow m$ scattering processes. For the self-similar distribution in eq.~(\ref{eq:scalingf}),
the scaling behavior of the collision integral  $C[p_T,p_z,\tau;f] = (Q \tau)^{\mu}\, C[(Q \tau)^{\beta}p_T,(Q \tau)^{\gamma}p_z;f_S]$, is described in terms of the exponent $\mu=\mu(\alpha,\beta,\gamma)$, the precise form of which depends on the underlying interaction. Substituting this scaling function into Eq.~(\ref{eq:Boltzmann}) leads to the {\it time-independent condition}
\begin{eqnarray}
\label{eq:attractor}
\alpha f_S(p_T,p_z)
+\beta p_T \partial_{p_T} f_S(p_T,p_z)  +  \left(\gamma-1\right) p_z \partial_{p_z} f_S(p_T,p_z)
= Q^{-1} C[p_T,p_z;f_S]\;. 
\end{eqnarray}
The non-thermal attractor solution observed in our lattice simulations is a nontrivial solution of this kinetic equation.

Such scaling analyses of kinetic equations, demonstrate that the scaling exponents characterizing the attractor can be classified on very general grounds of dimensionality, conservation laws and boundary conditions for the evolution~\cite{Micha:2004bv,Berges:2008wm}.  The observation of such a self-similar scaling solution of the expanding non-Abelian plasma is a powerful indication for universal behavior far from equilibrium, as seen in dynamical systems ranging from the inflaton dynamics of relativistic scalar fields to table-top cold atom experiments~\cite{Berges:2008wm,Berges:2008sr,Nowak:2011sk}. Thus while the bottom-up scenario correctly captures the universal properties of the turbulent thermalization process, the phenomenon may be more general than indicated by the specifics of the bottom-up scenario. Preliminary results from simulations of longitudinally expanding {\it scalar} fields appear to bolster these conclusions~\cite{Kirill}.

What then of isotropization and thermalization? Though the system does not free-stream and is strongly interacting, it does not isotropize either in the classical regime where the simulations are valid. In the bottom-up kinetic scenario, which matches our results in the classical regime, the system can be followed all the way to thermalization. In the quantum regime (where $f<1$), 
$2\leftrightarrow 3$ inelastic interactions become important.  For  $\alpha_S^{-3/2} < Q\tau < \alpha_S^{-5/2}$, soft gluons produced through inelastic $2\leftrightarrow 3$ interactions contribute very little to the total number but begin to dominate Debye screening. The latter begins to influence dynamics. In this regime, 
one finds that $\Lambda_L^2\sim \alpha_S Q^2$, independent of time, and $m_D^2 = \alpha_S^{3/4} Q^2/(Q\tau)^{1/2}$. The transverse pressure is still dominated by the hard gluons, and continues to have the parametric form $P_T\sim \alpha_S^{-1} Q^4/(Q\tau)$; the longitudinal pressure, on the other hand, is given by $P_L\sim \Lambda_L^2 m_D^2$. The ratio of the two then gives $P_L/P_T \sim (Q\tau)^{1/2}$--it thus begins to rise again in the quantum regime\footnote{I am indebted to Soeren Schlichting for helping clarify this point.}. 

\begin{figure}[t!]	
\vspace*{-3.5ex}
\includegraphics[width=0.55\textwidth]{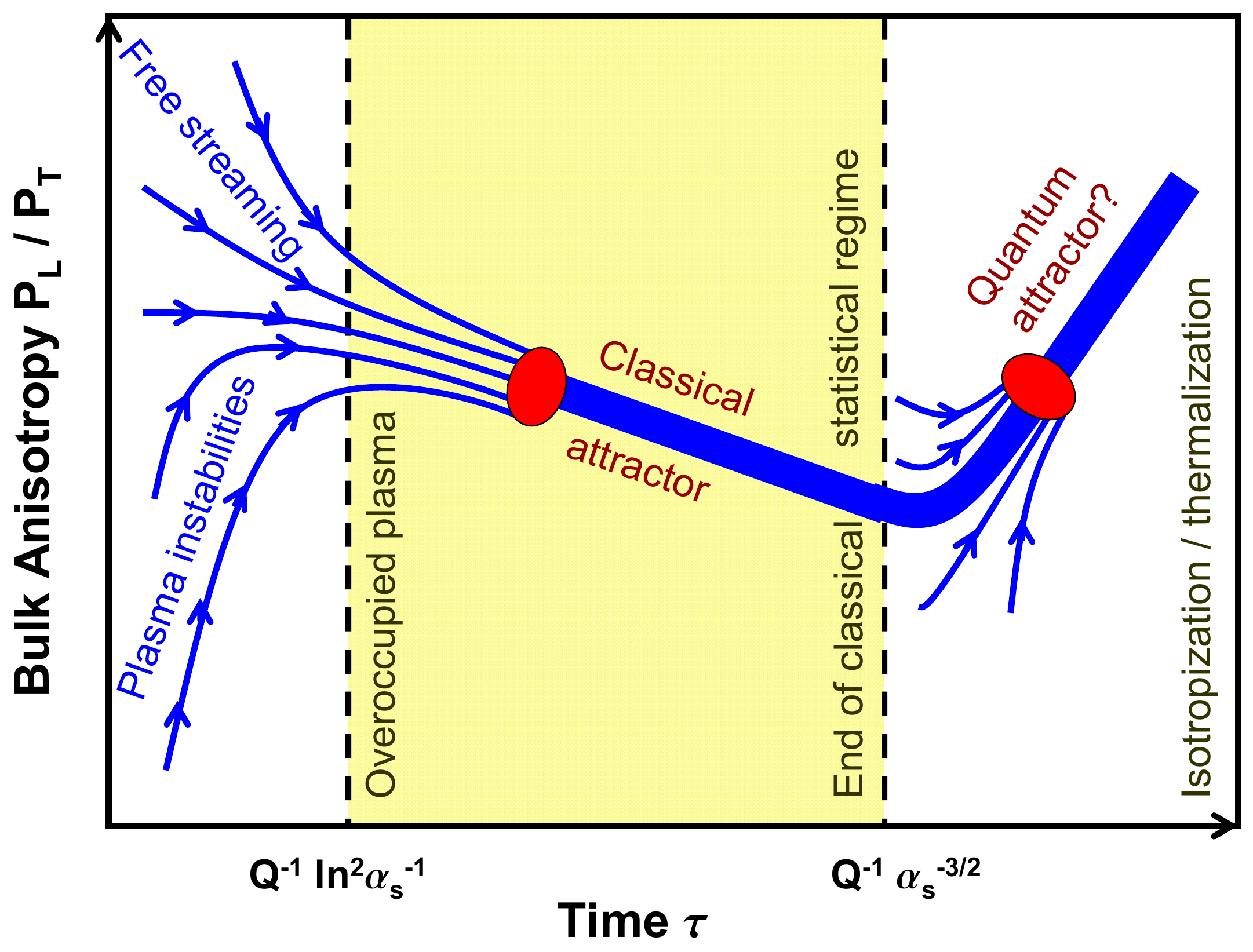}	
\caption{\label{fig:Summary} Schematic illustration of the thermalization process for the example of the bulk anisotropy. The time evolution in the shaded yellow regime pictures the results of this work, with the red ellipse symbolizing the observed non-thermal fixed point. The curves beyond the classical regime reflect the evolution in the bottom-up scenario \cite{bib:BMSS}, with the possibility of a second turbulent attractor in the quantum regime~\cite{Blaizot:2013hx}. From Ref.~\cite{Berges:2013eia}. } 
\end{figure}

By $Q\tau \sim \alpha_S^{-5/2}$, the number of soft gluons exceeds the number of hard gluons. The soft gluons collide frequently--the rate of expansion for $Q\tau > \alpha_S^{-5/2}$ is slower than the relaxation time of soft gluons. The soft sector has therefore thermalized by $Q\tau\sim \alpha_S^{-5/2}$. The system is however anistropic as a whole because of the remaining gluons in the hard sector.  It has recently been conjectured in the context of jet quenching that hard gluons in the quark-gluon plasma lose their energy via a medium induced  self-similar turbulent cascade of soft gluons~\cite{Blaizot:2013hx}--a quantum turbulent attractor. This framework should also be applicable to describe how the remaining hard gluons at the scale $Q$ are finally quenched and become part of the thermalized plasma. In the BMSS scenario, this occurs at the time $Q\tau = \alpha_S^{-13/5}$ and at a temperature of $T=\alpha_S^{1/2} Q$.  Subsequently, the temperature of the system decreases as $\tau^{-1/3}$ as for a hydrodynamic system undergoing one dimensional expansion. Fig.~(\ref{fig:Summary}) illustrates, via the temporal evolution of $P_L/P_T$, one scenario, based on classical-statistical field theory and kinetic theory, of the rich non-equilibrium structure that precedes the formation of a thermal quark-gluon plasma. 

While all the estimates here were obtained in a weak coupling framework, there is no show stopper for extending these ideas to larger, more realistic, couplings. If one puts in realistic values of $\alpha_S$, one gets reasonable ball park numbers for the saturation scales estimated at RHIC and LHC.  More quantitative phenomenology will depend on computing the pre-factors in these estimates; such computations are in progress~\cite{Kurkela}. Even though the system may not be describable by hydrodynamics at early times, it is still strongly interacting and generating flow. Hence it is plausible that when such effects are included, the start time for hydrodynamics could be moved to later times, and still give good agreement with flow data from RHIC and LHC. 
It is also interesting to ask if there are direct signatures in the data of anisotropic early time dynamics. Recent efforts have focused on electromagnetic signatures, in particular the photon spectrum~\cite{McLerran:2014hza}. 

An alternative scenario is implied by the numerical simulations in \cite{Gelis:2013rba} based on Eq.~(\ref{eq:quantum}). Recall this  expression is derived at $\tau_0=0^+$, in the theoretical formulation presented in \cite{Epelbaum:2013waa}. It is found that for 
$\alpha_S \sim 1/50$ ($g=0.5$), the ratio $P_L/P_T$ rises rapidly and appears to reach a constant value that is approximately 0.6 by $Q_S\tau=10$ and remains constant up to the maximum time presented in the simulation of $Q_S\tau=40$. On the surface, this appears to contradict the results of \cite{Berges:2013eia,Berges:2013fga} since in the latter $P_L/P_T$ decreases as $(Q\tau)^{-2/3}$. Assuming both numerical simulations are accurate, the difference could be attributed to a) the different couplings chosen in the  two simulations, and b) the different initial conditions corresponding to Eq.~(\ref{eq:quantum}) for \cite{Gelis:2013rba} and Eq.~(\ref{eq:quantum2}) for \cite{Berges:2013eia,Berges:2013fga}. 

The first point can be addressed straightforwardly. Simulations performed at $Q\tau_0=20$ (corresponding to $\alpha_S\sim 10^{-2}$), show the same late time behavior as for the larger values of $Q\tau_0$ shown in \cite{Berges:2013eia,Berges:2013fga}. A more contentious point could be the differing initial conditions. Our claim however is that they will give the same result at late times for both sets of initial conditions. To appreciate this, consider what the lifetime of the classical regime is for $\alpha_S = 1/50$, the value of the coupling in the simulations of \cite{Gelis:2013rba}. 
In the bottom-up scenario, this lifetime is $Q\tau_{\rm class.} = \alpha_S^{-3/2} \sim 350$. Other kinetic scenarios will give comparable times, if not longer, for the applicability of classical-statistical dynamics. The simulations presented in \cite{Gelis:2013rba} end at $Q\tau=40$, about $1/9$ of the lifetime of the classical regime. Our prediction is that if the simulations are followed for longer\footnote{At present, the simulations cannot be followed for longer times for technical reasons~\cite{Epelbaum}.}, within the time when classical dynamics is valid, the ratio of $P_L/P_T$ will turn over and decrease as a function of $\tau$. 
Another hint that this may be the right interpretation of the results of \cite{Gelis:2013rba}, is the fact that the end of the instability regime is at $Q\tau= \ln^2{\alpha_S^{-1}} \sim 15$, which is roughly the time when $P_L/P_T$ becomes large in the simulations of \cite{Gelis:2013rba}.  Finally, simulations of a scalar field theory at weak coupling with i) the `field driven' initial conditions of Eq.~(\ref{eq:quantum}) and ii) the `fluctuation driven' initial conditions of Eq.~(\ref{eq:quantum2}), give the same late time attractor behavior, well within the domain of validity of classical statistical simulations~\cite{Berges:2013lsa}.

The situation can be settled concretely by performing the simulations of \cite{Gelis:2013rba} on larger lattices for later times. 
It will also be important to extract the quasi-particle spectrum of \cite{Gelis:2013rba}, as for instance, done in \cite{Berges:2013eia,Berges:2013fga}. It is of crucial importance to settle this issue one way or the other. If the claim of \cite{Gelis:2013rba} is borne out, classical field dynamics can be matched to viscous hydrodynamics, in a more sophisticated version of the IP-Glasma model~\cite{STV}. If not, one has to work harder, and match classical dynamics to kinetic theory first, before thermalization is attained. It is at one level more cumbersome, but on another, provides further insight into the rich dynamics that becomes possible in strongly correlated non-Abelian plasmas. 

Interesting ideas to merge classical field theory and kinetic theory can be found in \cite{Attems:2012js,York:2014wja}. 
Of course, weak coupling approaches could be wrong altogether, and nature, at realistic coupling, may have different plans in mind. The data from heavy ion collisions however is so plentiful, and of such high quality, that one can expect significant progress in the near future. One line of practical inquiry is to see how tweaks of the early time dynamics in the very successful IP-Glasma model influence comparisons with experiment.

\section{The ridge in p+p, p+A and A+A collisions: the smallest thermal fluids?}

Data  on two particle correlations in high multiplicity proton-proton and proton-nucleus collisions at the LHC have a component 
that is only weakly dependent on the rapidity separation ($\Delta\eta$) between the pairs and is strongly collimated in the 
azimuthal angular separation ($\Delta\Phi\sim 0$) between the pairs. Such ``ridge"-like structures were first seen in A+A collisions at RHIC, at the LHC, and more recently, in very central deuteron-gold collisions at RHIC. For recent reviews, see \cite{Li:2012hc,Loizides:2013nka,Wang:2013qca}. 

\subsection{Initial state contributions to the ridge}
In perturbative QCD the only two parton correlation that one obtains is the back-to-back $\Delta\Phi\sim \pi$ correlation from the di-jet graph; this explains why none of the event generators saw the ridge collimation at $\Delta\Phi\sim 0$. In the CGC effective theory, the nearside $\Delta\Phi\sim 0$ collimation is obtained from connected two gluon production QCD graphs called ``Glasma graphs". They are QCD interference graphs. In conventional perturbative QCD computations, these graphs are ignored for good reason. Their contribution at high $p_T$ and in peripheral collisions is negligibly small. 

However,  most remarkably, the high occupancy of gluons (for transverse momenta $k_\perp \leq Q_S$, where $Q_S$ is the saturation scale) in rare high multiplicity proton-proton events enhances such graphs by $\alpha_S^{-8}$. This corresponds to a strikingly large enhancement of $\sim 10^5$ for typical values of the probed QCD fine structure constant $\alpha_S$! Thus in the CGC power counting, gluon saturation ensures that Glasma graphs provide a significant  additional contribution in high multiplicity events to ``di-jet" QCD graphs. The importance of Glasma graphs was first discussed in ~\cite{Dumitru:2008wn} and the formalism developed in ~\cite{Gelis:2008sz,Dusling:2009ni}. It was first postulated as an explanation of the high multiplicity CMS proton-proton ridge in \cite{Dumitru:2010iy}, and a quantitative description of the nearside collimated yield obtained in \cite{Dusling:2012iga}. 

The di-jet contribution that is long range in rapidity is described by BFKL dynamics~\cite{Balitsky:1978ic,Kuraev:1977fs}. We showed in \cite{Dusling:2012cg} that BFKL dynamics does well in describing the awayside spectra in high multiplicity proton-proton collisions. The description is significantly better than PYTHIA-8~\cite{Khachatryan:2010gv}, and $2\rightarrow 4$ QCD graphs in the Quasi--Multi--Regge--Kinematics (QMRK)~\cite{Fadin:1996zv,Leonidov:1999nc}. Both of these approaches overestimate the awayside yield especially at larger momenta. 

The Glasma+BFKL CGC framework~\cite{Dusling:2013oia,Dusling:2012wy} describes reasonably well the associated yield per trigger obtained in p+Pb data at $\sqrt{s}=5.02$ TeV/nucleon by the CMS collaboration~\cite{CMS:2012qk}. Subsequently, both the ALICE~\cite{Abelev:2012ola} and ATLAS~\cite{Aad:2012gla} collaborations published their di-hadron correlation results from the first LHC p+Pb run. The ALICE experiment has an acceptance in $\Delta \eta$ of $|\Delta \eta|<1.8$, while the ATLAS experiment has an acceptance of $2 < |\Delta \eta|< 5$, close to the CMS acceptance of $2<|\Delta \eta|<4$.  In addition to the LHC results, the PHENIX collaboration at RHIC reanalyzed their deuteron-gold data at 200 GeV/nucleon and extracted a ridge signal in very central events~\cite{Adare:2013piz}.  All three of the experiments show that when the two particle yield in peripheral collisions is subtracted from the central events, a dipole structure remains that is long range in rapidity.  This is precisely what  we anticipated in our Glasma+BFKL graph scenario. Specifically, this is because the BFKL di-jet contribution has a weak dependence on centrality and the net Glasma graph contribution is symmetric around $\Delta\phi = \pi/2$. 

The LHC experiments have analyzed~\cite{Abelev:2012ola,Aad:2012gla,Chatrchyan:2013nka} the $v_n$ Fourier moments of the jet-subtracted di-hadron yields in proton-nucleus collisions. (Note that this has not been done yet in proton-proton high multiplicity collisions because the jet yield is very large, potentially generating large systematic uncertainties in any such subtraction.) They obtained a significant value of $v_2$ which approaches that observed in peripheral A+A collisions. Because of the $\Delta\Phi$ dependence of our Glasma graph contribution, we too expect a significant $v_2$. However, the experiments observed a distinct $v_3$ contribution. In addition, $v_2\{4\}$, $v_2$ extracted from four particle cumulants, is significant. Both of these effects were widely interpreted as problematic for the initial state CGC picture. However, there is also an interference contribution between the two contributions that must be there. Previously, we had not included it because it is small. Our preliminary conclusions are that this contribution gives a triangular azimuthal anisotropy and that it has the right centrality and $p_T$ dependence seen in the data~\cite{Kevin}. 

\subsection{Final state effects and the ridge}

The observation of large ridges in proton-nucleus collisions that are comparable in size to those in peripheral nucleus-nucleus collisions have reinvigorated interpretations in terms of hydrodynamic flow.  Additional force for the flow argument is adduced from the mass ordering of $v_2(p_T)$ for different hadron species~\cite{ABELEV:2013wsa}. In p+A collisions, it looks remarkably similar to that in peripheral A+A collisions, both in shape and magnitude. Computations in a hydrodynamical model reproduced a number of features of the proton-lead data~\cite{Bozek:2012gr,Bozek:2013df,Bozek:2013uha,Bozek:2013ska}. Other recent hydrodynamics based discussions of the ridge in p+A collisions can be found in refs.~\cite{Shuryak:2013ke,Qin:2013bha,Werner:2013ipa}. However, hydrodynamics when applied to p+A collisions is very sensitive to assumptions about the dynamics of the initial state and values of $\eta/s$.  

The hydrodynamic description of small size systems has recently been studied in the framework of the IP-Glasma+MUSIC model ~\cite{STV}.  Though the model gives excellent fits to data at both RHIC and LHC in A+A collisions, the values of $v_2$ and $v_3$ in $p+A$ collisions are factors of 2-3 smaller than the data for the same values of $\eta/s$ that describe the A+A data~\cite{Bzdak:2013zma}. To help distinguish between final state models alone, it is essential to perform `apples-to-apples' comparisons of  data in very different centrality classes in A+A and p+A that correspond to the same $N_{\rm charge}$. Very recently, we performed a comprehensive study of single inclusive multiplicities and multiplicity distributions in p+p, p/d+A and A+A collisions within the IP-Glasma framework~\cite{Schenke:2013dpa}. With the results of this study, flow generated from the initial state configurations that generate the $70$\% central A+A collisions can be compared directly to the flow generated from initial state configurations corresponding to $3$\% central events in p+Pb collisions. The preliminary results are consistent with the studies in \cite{Bzdak:2013zma} that indicate that $v_2$ generated is too small to describe the data. We note however that if the spatial distribution of gluons is less uniform than that given by the IP-Glasma framework, larger flow moments may be generated. This further emphasizes the sensitivity of results to assumptions 
about the spatial distribution of glue in rare configurations of the colliding protons and nuclei. 

\subsection{The ridge through my colored glasses}

A detailed discussion of our take  on the initial state versus final state issues on the ridge in p+p, p+A and A+A collisions can be found in \cite{Venugopalan:2013cga}. Here we will provide a rough sketch of our perspective on these issues. In high multiplicity events, the power counting in the CGC tells us that the underlying Glasma background is azimuthally isotropic. This contribution is the leading contribution in powers of the coupling and the number of colors $N_c$. Jetty contributions are sub-leading in powers of the coupling but leading in $N_c$. The interplay between the two, as a function of centrality and transverse momenta, determines the broad features of two particle correlations in high multiplicity events. Similar conclusions, couched in the language of ``sphericity" observables, have been advanced in the PYTHIA event generator framework~\cite{Cuautle:2014yda}.

However, in addition to the leading azimuthally isotropic and jetty contributions in the Glasma, there is an intrinsic collimation in Delta $\Phi$ that is long range in rapidity. These arise from the so-called Glasma graphs we described previously. They are leading in powers of the coupling in high multiplicity events but are sub-leading ($1/N_c^2$ suppressed) in the number of colors. The effect is purely a quantum interference effect, because the color structure of the graphs in the amplitude and in the complex conjugate amplitude are different. 

Some of the Glasma contributions are Hanbury-Brown--Twiss (HBT) like interference contributions, but others are not--the collimation is sensitive to the detailed shape of the unintegrated gluon distributions in the protons and nuclei. In A+A collisions, the azimuthally isotropic leading contributions to multiparticle production develop a sensitivity to the geometry of the overlap region due to final state interactions and generate the long range ridge like structure due to flow. The Glasma graphs are tiny in comparison. However, when the multiplicity of the isotropic contribution is small, the intrinsic Glasma contribution has the chance to peek its head out of the underlying background, a colorful diamond shining through the muck. This long range quantum correlation provides an intriguing alternative to the ubiquitous flow explanation that is applied to the small sized p+A and p+p systems. When we put together the following 
\begin{itemize}
\item  Our recent results on the triangular anisotropy in this framework, 
\item  the disagreement in the IP-Glasma+MUSIC model with $v_2$ and $v_3$  measured in p+A collisions, especially when contrasted to the remarkable success of the IP-Glasma+MUSIC model in describing A+A collisions, 
\item the recent HBT results from the ALICE collaboration showing that HBT radii in p+A collisions 
have the same slope in p+p and p+A collisions, in contrast to A+A collisions~\cite{Abelev:2014pja},
\end{itemize}
there is some cause for optimism that this manifestation of long range gluon `entanglement' will survive. In addition to further developments in theory, a wealth of data coming out in the near future will further test this picture relative to other, mostly final state, explanations. 

\section{G.E.B: Sui generis, requiescat in pace}

I first met Gerry Brown when he invited me to visit Stony Brook, whilst I was still an undergraduate at the University of Chicago. He asked me if I had the proverbial fire in my belly. I mumbled something, and that was that--I was hired. It being 1987, Stony Brook was the tower of BBAL~\cite{Bethe:1979zd}, with Gerry conducting that noisy and unruly orchestra with aplomb. The wild excitement notwithstanding, supernova physics was at that time not quite to my taste. But Gerry didn't hold it against me, and instead paired me with Prakash, who proved to be a model advisor. Gerry persisted in being very kind to me and sent me off to spend summers in Berkeley, Copenhagen and Minneapolis. I admired greatly Gerry's intuitive approach and passion for physics that he communicated effortlessly, as well as his wide erudition and dry wit. I was fortunate though not to catch the sharp edge of the last, which did not endear him to everyone. When I was about to graduate, and job prospects looked dim, Gerry came to me and said I should stay an extra year and work with him if things didn't work out. They did work out but part of me regrets that missed opportunity. 

\section*{Acknowledgments}
R.V's research was supported by DOE Contract No.
DE-AC02-98CH10886. He thanks his collaborators Juergen Berges, Kirill Boguslavski, Adam Bzdak, Kevin Dusling, Bjoern Schenke, Soeren Schlichting and Prithwish Tribedy for generously contributing their insight into the topics discussed here. He is grateful to Bjoern Schenke and Soeren Schlichting for a close reading of the manuscript, and to Edward Shuryak for valuable editorial comments. 






\begin{thebibliography}{00}


\bibitem{Arnold:1997gh} 
  P.~B.~Arnold,  L.~G.~Yaffe,
  Phys.\ Rev.\ D {\bf 57}, 1178 (1998).

\bibitem{STV}B.~Schenke, P.~Tribedy, R.~Venugopalan,
  Phys.\ Rev.\ Lett.\  {\bf 108}, 252301 (2012); Phys.\ Rev.\ C {\bf 86}, 034908 (2012); C.~Gale, S.~Jeon, B.~Schenke, P.~Tribedy, R.~Venugopalan,
Phys.\ Rev.\ Lett.\  {\bf 110}, 012302 (2013).

\bibitem{Gribov:1984tu} 
  L.~V.~Gribov, E.~M.~Levin and M.~G.~Ryskin,
  Phys.\ Rept.\  {\bf 100}, 1 (1983).

\bibitem{Mueller:1985wy} 
  A.~H.~Mueller and J.~-w.~Qiu,
  Nucl.\ Phys.\ B {\bf 268}, 427 (1986). 

\bibitem{CGC}
F. Gelis, E. Iancu, J. Jalilian-Marian, R. Venugopalan, Ann. Rev. Part. Nucl.
  Sci. {\bf 60}, 463 (2010).

\bibitem{MV}L.D. McLerran, R. Venugopalan,  Phys. Rev. {\bf D} {\bf 49}, 2233 (1994); {\it ibid.} {\bf 49}, 3352 (1994); {\it ibid.} {\bf 50}, 2225 (1994).

\bibitem{CYM}
  A.~Kovner, L.~D.~McLerran, H.~Weigert,
  Phys.\ Rev.\  {\bf D52}, 3809-3814 (1995).

\bibitem{CYM-numerics} 
A.~Krasnitz, R.~Venugopalan,
  Nucl.\ Phys.\  {\bf B557}, 237 (1999).

\bibitem{Glasma}
  T.~Lappi, L.~McLerran,
  Nucl.\ Phys.\  {\bf A772}, 200-212 (2006); F.~Gelis, R.~Venugopalan,
  Acta Phys.\ Polon.\ B {\bf 37}, 3253 (2006); T.~Lappi, 
  Int.\ J.\ Mod.\ Phys.\ E {\bf 20}, 1 (2011).

\bibitem{CYM-numerics2}
 A.~Krasnitz and R.~Venugopalan,
  Phys.\ Rev.\ Lett.\  {\bf 86}, 1717 (2001); A.~Krasnitz, Y.~Nara and R.~Venugopalan,
  Phys.\ Rev.\ Lett.\  {\bf 87}, 192302 (2001);  T.~Lappi,
  Phys.\ Rev.\  {\bf C67}, 054903 (2003).

\bibitem{Weibel} 
  S.~Mrowczynski,
  Phys.\ Lett.\ B {\bf 314}, 118 (1993); P.~Romatschke and M.~Strickland,
  Phys.\ Rev.\ D {\bf 68}, 036004 (2003);  P.~B.~Arnold, J.~Lenaghan and G.~D.~Moore,
  JHEP {\bf 0308}, 002 (2003).

\bibitem{Nielsen-Olesen}H.~Fujii, K.~Itakura, A.~Iwazaki,
  Nucl.\ Phys.\ A {\bf 828}, 178 (2009)
  [arXiv:0903.2930 [hep-ph]].

\bibitem{Romatschke:2005pm}
  P.~Romatschke, R.~Venugopalan,
  Phys.\ Rev.\ Lett.\  {\bf 96}, 062302 (2006); 
  Phys.\ Rev.\  D {\bf 74}, 045011 (2006);  Eur.\ Phys.\ J.\  {\bf A29}, 71-75 (2006); 
 K.~Fukushima, F.~Gelis, Nucl.\ Phys.\ A {\bf 874}, 108 (2012).

\bibitem{Gelis:2007kn} 
  F.~Gelis, T.~Lappi and R.~Venugopalan,
  Int.\ J.\ Mod.\ Phys.\ E {\bf 16}, 2595 (2007).

\bibitem{Dusling:2011rz}
{K. Dusling, F. Gelis, R. Venugopalan}, Nucl. Phys. {\bf A} {\bf 872}, 161
  (2011).

\bibitem{Epelbaum:2013waa} 
  T.~Epelbaum and F.~Gelis,
  Phys.\ Rev.\ D {\bf 88}, 085015 (2013).

\bibitem{Berges:2004ce} 
  J.~Berges, S.~Borsanyi and C.~Wetterich,
  Phys.\ Rev.\ Lett.\  {\bf 93}, 142002 (2004).

\bibitem{Dusling:2010rm} 
  K.~Dusling, T.~Epelbaum, F.~Gelis and R.~Venugopalan,
  Nucl.\ Phys.\ A {\bf 850}, 69 (2011).

\bibitem{Berges:2013eia} 
  J.~Berges, K.~Boguslavski, S.~Schlichting and R.~Venugopalan,
  Phys.\ Rev.\ D {\bf 89}, 074011 (2014).

\bibitem{Berges:2013fga} 
  J.~Berges, K.~Boguslavski, S.~Schlichting and R.~Venugopalan,
  arXiv:1311.3005 [hep-ph].

\bibitem{Kurkela:2012hp} 
  A.~Kurkela and G.~D.~Moore,
  Phys.\ Rev.\ D {\bf 86}, 056008 (2012).

\bibitem{Mueller:2002gd} 
  A.~H.~Mueller and D.~T.~Son,
  Phys.\ Lett.\ B {\bf 582}, 279 (2004).

\bibitem{Jeon:2004dh} 
  S.~Jeon,
  Phys.\ Rev.\ C {\bf 72}, 014907 (2005).

\bibitem{bib:KMII}
  A.~Kurkela and G.~D.~Moore,
  JHEP {\bf 1111}, 120 (2011).

\bibitem{bib:BGLMV}
 J.~-P.~Blaizot, F.~Gelis, J.~-F.~Liao, L.~McLerran and R.~Venugopalan,
  Nucl.\ Phys.\ A {\bf 873}, 68 (2012).

\bibitem{bib:BMSS}
R.~Baier, A.~H.~Mueller, D.~Schiff and D.~T.~Son,
  Phys.\ Lett.\ B {\bf 502}, 51 (2001).

\bibitem{Micha:2004bv} 
  R.~Micha and I.~I.~Tkachev,
  Phys.\ Rev.\ D {\bf 70}, 043538 (2004). 

\bibitem{Berges:2008wm} 
  J.~Berges, A.~Rothkopf and J.~Schmidt,
  Phys.\ Rev.\ Lett.\  {\bf 101}, 041603 (2008). 

\bibitem{Berges:2008sr} 
  J.~Berges and G.~Hoffmeister,
  Nucl.\ Phys.\ B {\bf 813}, 383 (2009).

\bibitem{Nowak:2011sk} 
  B.~Nowak, J.~Schole, D.~Sexty and T.~Gasenzer,
  Phys.\ Rev.\ A {\bf 85}, 043627 (2012).

\bibitem{Kirill}K. Boguslavski, talk at RBRC workshop on ``The approach to equilibration in strongly interacting matter", 
BNL, April 2-4, 2014.

\bibitem{Kurkela}A. Kurkela, work in progress, private communication. 

\bibitem{Blaizot:2013hx} 
  J.~-P.~Blaizot, E.~Iancu and Y.~Mehtar-Tani,
  Phys.\ Rev.\ Lett.\  {\bf 111}, 052001 (2013).

\bibitem{McLerran:2014hza} 
  L.~McLerran and B.~Schenke,
  arXiv:1403.7462 [hep-ph].

\bibitem{Gelis:2013rba} 
  T.~Epelbaum and F.~Gelis,
  Phys.\ Rev.\ Lett.\  {\bf 111}, 232301 (2013). 

\bibitem{Epelbaum}T. Epelbaum and F. Gelis, private communication. 

\bibitem{Berges:2013lsa} 
  J.~Berges, K.~Boguslavski, S.~Schlichting and R.~Venugopalan,
  arXiv:1312.5216 [hep-ph].

\bibitem{Attems:2012js} 
  M.~Attems, A.~Rebhan and M.~Strickland,
  Phys.\ Rev.\ D {\bf 87}, 025010 (2013).

\bibitem{York:2014wja} 
  M.~C.~A.~York, A.~Kurkela, E.~Lu and G.~D.~Moore,
  arXiv:1401.3751 [hep-ph].

\bibitem{Li:2012hc} 
  W.~Li,
  Mod.\ Phys.\ Lett.\ A {\bf 27}, 1230018 (2012).

\bibitem{Loizides:2013nka} 
  C.~Loizides,
  arXiv:1308.1377 [nucl-ex].

\bibitem{Wang:2013qca} 
  F.~Wang,
  Prog.\  Part.\  Nucl.\  Phys.\  {\bf 74}, 35 (2014).


\bibitem{Dumitru:2008wn} 
  A.~Dumitru, F.~Gelis, L.~McLerran and R.~Venugopalan,
  Nucl.\ Phys.\ A {\bf 810}, 91 (2008). 

\bibitem{Gelis:2008sz} 
  F.~Gelis, T.~Lappi and R.~Venugopalan,
  Phys.\ Rev.\ D {\bf 79}, 094017 (2009).

\bibitem{Dusling:2009ni} 
  K.~Dusling, F.~Gelis, T.~Lappi and R.~Venugopalan,
  Nucl.\ Phys.\ A {\bf 836}, 159 (2010).

\bibitem{Dumitru:2010iy} 
  A.~Dumitru, K.~Dusling, F.~Gelis, J.~Jalilian-Marian, T.~Lappi and R.~Venugopalan,
  Phys.\ Lett.\ B {\bf 697}, 21 (2011).

\bibitem{Dusling:2012iga} 
  K.~Dusling and R.~Venugopalan,
  Phys.\ Rev.\ Lett.\  {\bf 108}, 262001 (2012).

\bibitem{Dusling:2012cg} 
  K.~Dusling and R.~Venugopalan,
  Phys.\ Rev.\ D {\bf 87}, no. 5, 051502 (2013).

\bibitem{Balitsky:1978ic} 
  I.~I.~Balitsky and L.~N.~Lipatov,
  Sov.\ J.\ Nucl.\ Phys.\  {\bf 28}, 822 (1978)
  [Yad.\ Fiz.\  {\bf 28}, 1597 (1978)].

\bibitem{Kuraev:1977fs} 
  E.~A.~Kuraev, L.~N.~Lipatov and V.~S.~Fadin,
  Sov.\ Phys.\ JETP {\bf 45}, 199 (1977).

\bibitem{Fadin:1996zv} 
  V.~S.~Fadin, M.~I.~Kotsky and L.~N.~Lipatov,
  hep-ph/9704267.

\bibitem{Leonidov:1999nc} 
  A.~Leonidov and D.~Ostrovsky,
  Phys.\ Rev.\ D {\bf 62}, 094009 (2000)
  [hep-ph/9905496].

\bibitem{Dusling:2013oia} 
  K.~Dusling and R.~Venugopalan,
  Phys.\ Rev.\ D {\bf 87}, 094034 (2013).

\bibitem{Dusling:2012wy} 
  K.~Dusling and R.~Venugopalan,
  Phys.\ Rev.\ D {\bf 87}, no. 5, 054014 (2013).

\bibitem{Khachatryan:2010gv} 
  V.~Khachatryan {\it et al.}  [CMS Collaboration],
  JHEP {\bf 1009}, 091 (2010).

\bibitem{CMS:2012qk} 
  S.~Chatrchyan {\it et al.}  [CMS Collaboration],
  Phys.\ Lett.\ B {\bf 718}, 795 (2013).

\bibitem{Aad:2012gla} 
  G.~Aad {\it et al.}  [ATLAS Collaboration],
  Phys.\ Rev.\ Lett.\  {\bf 110}, 182302 (2013).

\bibitem{Abelev:2012ola} 
  B.~Abelev {\it et al.}  [ALICE Collaboration],
  Phys.\ Lett.\ B {\bf 719}, 29 (2013).

\bibitem{Adare:2013piz}
  A.~Adare {\it et al.}  [PHENIX Collaboration],
  [arXiv:1303.1794 [nucl-ex]].

\bibitem{Chatrchyan:2013nka} 
  S.~Chatrchyan {\it et al.}  [CMS Collaboration],
  Phys.\ Lett.\ B {\bf 724}, 213 (2013).

\bibitem{Kevin}K. Dusling and R. Venugopalan, {\it in preparation}.

\bibitem{ABELEV:2013wsa} 
  B.~B.~Abelev {\it et al.}  [ALICE Collaboration],
  Phys.\ Lett.\ B {\bf 726}, 164 (2013).

\bibitem{Bozek:2012gr} 
  P.~Bozek and W.~Broniowski,
  Phys.\ Lett.\ B {\bf 718}, 1557 (2013).

\bibitem{Bozek:2013df} 
  P.~Bozek and W.~Broniowski,
  Phys.\ Lett.\ B {\bf 720}, 250 (2013)
  [arXiv:1301.3314 [nucl-th]].

\bibitem{Bozek:2013uha} 
  P.~Bozek and W.~Broniowski,
  Phys.\ Rev.\ C {\bf 88}, 014903 (2013).

\bibitem{Bozek:2013ska} 
  P.~Bozek, W.~Broniowski and G.~Torrieri,
  Phys.\ Rev.\ Lett.\  {\bf 111}, 172303 (2013).

\bibitem{Shuryak:2013ke} 
  E.~Shuryak and I.~Zahed,
  Phys.\ Rev.\ C {\bf 88}, 044915 (2013)
  [arXiv:1301.4470 [hep-ph]].

\bibitem{Qin:2013bha} 
  G.~-Y.~Qin and B.~Muller,
  arXiv:1306.3439 [nucl-th].

\bibitem{Werner:2013ipa} 
  K.~Werner, M.~Bleicher, B.~Guiot, I.~.Karpenko and T.~Pierog,
  arXiv:1307.4379 [nucl-th].

\bibitem{Schenke:2013dpa} 
  B.~Schenke, P.~Tribedy and R.~Venugopalan,
  Phys.\ Rev.\ C {\bf 89}, 024901 (2014).

\bibitem{Bzdak:2013zma} 
  A.~Bzdak, B.~Schenke, P.~Tribedy and R.~Venugopalan,
  Phys.\ Rev.\ C {\bf 87}, 064906 (2013).

\bibitem{Venugopalan:2013cga} 
  R.~Venugopalan,
  arXiv:1312.0113 [hep-ph].

\bibitem{Cuautle:2014yda} 
  E.~Cuautle, R.~Jimenez, I.~Maldonado, A.~Ortiz, G.~Paic and E.~Perez,
  arXiv:1404.2372 [hep-ph].

\bibitem{Abelev:2014pja} 
  B.~B.~Abelev {\it et al.}  [ALICE Collaboration],
  arXiv:1404.1194 [nucl-ex].

\bibitem{Bethe:1979zd} 
  H.~A.~Bethe, G.~E.~Brown, J.~Applegate and J.~M.~Lattimer,
  Nucl.\ Phys.\ A {\bf 324}, 487 (1979).




 \end{thebibliography}



\end{document}